\documentclass[prb,twocolumn,amsmath,amssymb,showpacs]{revtex4}
\usepackage{graphicx}

\begin{document}


\title{On the dielectric susceptibility calculation in the incommensurate
phase of $\bf\rm K_2SeO_4$.}

\author{T.A. Aslanyan}
\email{aslanyan@freenet.am}
\affiliation{Institute for Physical
Research, Armenian National Academy of Sciences, Ashtarak-2,
378410 Armenia}



\begin{abstract}
It is shown that the thermodynamic potential of the domain-like
incommensurate (IC) phase  of the $\bf\rm K_2SeO_4$ crystal
(viewed as a model for the IC-C transition) should be supplemented
with a term, taking into account the local, Lorentz electric
field. The latter qualitatively changes the result of calculation
of the dielectric susceptibility for this IC structure by
Nattermann and Trimper, J. Phys. C: Solid State Phys. \textbf{14},
1603, (1981), and gives phase transition to the ferroelectric IC
phase obtained by Aslanyan, Phys. Rev. B \textbf{70}, 024102,
(2004).
\end{abstract}
\maketitle

In the present paper the behavior of the incommensurate (IC) phase
of the $\bf\rm K_2SeO_4$ crystal near the lock-in transition is
discussed. We remind that $\bf\rm K_2SeO_4$ undergoes on cooling
two successive phase transitions: the IC transition at $T_i=129K$
and the lock-in transition transition to the triple-period
commensurate ferroelectric phase at $T_c=93K$.\cite{I,Iizumi}
Nattermann and Trimper \cite{Natter} calculated the dielectric
susceptibility $\chi$ of the crystal in the IC phase, and showed
that $\chi$ diverges on approaching the lock-in transition point
(i.e., a loss of stability of the IC phase with respect to the
commensurate phase takes place). The dielectric susceptibility for
the same IC structure was calculated also by
Aslanyan,\cite{Aslanyan} and it was obtained that $\chi$ diverges
not in the lock-in transition point, but at a higher temperature,
giving rise to the phase transition from the IC to the
ferroelectric IC structure. Spatial distributions of the IC
modulation phase and polarization vector in the ferroelectric (or
$z$- polarized) IC phase \cite{Aslanyan} are depicted in Fig.1.
Below we explain the origin of the difference between the results
of the two calculations, \cite{Natter,Aslanyan} and show that the
thermodynamic potential used for calculations by Nattermann and
Trimper \cite{Natter} should be supplemented with a term, taking
into account the local (Lorentz) microscopic electric field,
induced by the dipoles (the neighboring pairs of the domains in
Fig.1 can be viewed as periodically arranged dipoles). In such a
case, calculation of the dielectric susceptibility by Nattermann
and Trimper \cite{Natter} should be revised, and it reduces to
that carried out by Aslanyan.\cite{Aslanyan}

The starting point in both calculations \cite{Natter,Aslanyan} is
the thermodynamic potential of the following form:
\begin{eqnarray}
\nonumber &&\tilde\Phi =\int dx\{
\frac{D}{2}\eta_0^2(\frac{\partial\varphi} {\partial x})^2
+\frac{f}{2}\eta_0^6\cos (6kx+6\varphi )+\\
&&\frac{r}{2}P\eta_0^3\cos (3kx+ 3\varphi )
+\frac{\chi^{-1}_0}{2}P^2 \}
\end{eqnarray}
In this expression (and below as well) the notations of paper [4]
are used, where $\eta_0$ and $\varphi$  are, respectively, the
amplitude and the phase of the IC modulation, $k$ is the IC vector
at the temperature of the IC transition (from the normal to the IC
phase), $P$ is the polarization vector in the $z$ direction, and
the IC modulation is given by $\eta_0\cos (kx+ \varphi )$.
\begin{figure}[h]
\mbox{\includegraphics[scale=0.9]{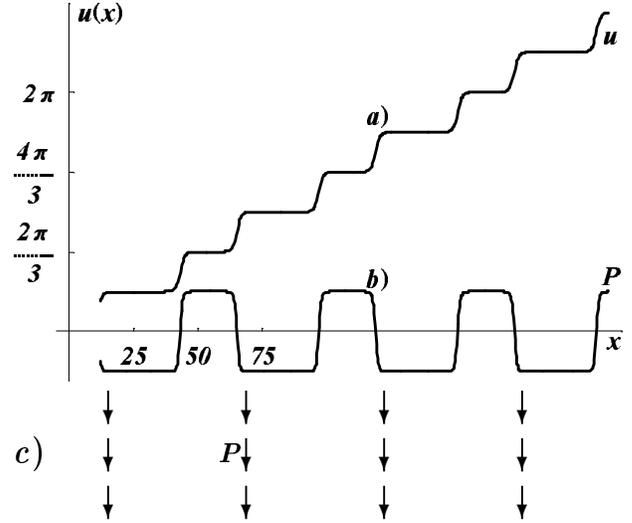}}
 \caption{a) Coordinate $x$
dependence of the $z$ -polarized IC structure's modulation phase
$u(x)=\varphi (x) +kx$. Each plateau corresponds to the domain
with + or - polarization vector $P$, which is along the $\pm z$
direction (curve b). Since the sizes of the adjacent domains are
different, the integral polarization of any two adjacent domains
(and, consequently, of the crystal as a whole) is not zero.
Polarization of the crystal appears due to the displacement of the
odd domain walls with respect to the even walls, which makes
nonequal the sizes of adjacent domains. The coordinate $x$ is
given in multiples of the lattice parameter. c) Periodic
disposition of the dipoles in the $x-z$ plane. These dipoles
correspond to the pairs of adjacent domains shown by the curve b).
In the $y$ and $z$ directions the periods are equal to the crystal
lattice parameters, while in the $x$ direction the distance
between dipoles equals to the width of the pair of domains.}
\end{figure}

However, potential (1) should be supplemented with a term of the
following origin. The domain-like IC structure, which develops
from the sinusoidal one on cooling, can be polarized in the $z$
direction via displacements of the odd domain walls with respect
to the even walls in the $x$ direction, as shown in Fig.1. The
pairs of adjacent domains (which become non-equisize after such
displacements) can be viewed as a periodically arranged dipoles as
in Fig.1, c). However, any periodically (or chaotically, as in the
case of a gas) distributed dipoles, with nonzero integral
polarization, induce a microscopic electric field $E_{micr}$ (the
so called local Lorentz field) biassing each dipole. For the case
of a gas or a cubic crystal, this field is equal to $E_{micr}=4\pi
\bar P/3$, where $\bar P$ is the spatially averaged polarization.
The effect of the microscopic electric field is implicitly taken
into account for the susceptibility $\chi_0$ of the high
temperature phase. However, in the IC phase, when the crystal
acquires a new sublattice in form of the polarizing pairs of
domains, one should additionally take it explicitly into account.
So, for the orthorhombic domain lattice of $\bf\rm K_2SeO_4$ the
microscopic field is
$$E_{micr}= a\bar P, \,\; {\rm with}\; \bar P= \frac{1}{L}\int
P(R)dx,$$ where $a$ is a coefficient of the same order as $4\pi
/3$ (in case of a cubic symmetry $a=4\pi /3$), and $L$ is the
crystal's length in the $x$ direction. Due to the appearance of
the field, biassing each dipole, the thermodynamic potential (1)
should be contributed by the additional term $$-\bar
PE_{micr}=-a\bar P^2=-a\left( \int P(R)dx \right)^2/L^2,$$ which
should be added to the expression under integral (1). It should be
emphasized a non-local character of this energy term (square of
the integral), which significantly changes the expected results of
calculations.

In what follows we show that the thermodynamic potential of the
crystal with the domain-induced microscopic electric field can be
reduced to the potential, analyzed in the paper [4], and the
behavior of the system in such a case should be that described
ibidem. Minimizing the potential over $P(R)$ and $\bar P$, under
the condition
$$\int (P(r)-\bar P)dx=0,$$ which should be added to the potential
equation with the Lagrange multiplier, one obtains
\begin{eqnarray*}
P(R)=  - \frac{r\chi_{0}\eta_0^3}{2}\cos 3u -
\frac{ar\chi_{0}^2\eta_0^3}{L(1-2\chi_0 a)}\int \cos 3u\, dx
\end{eqnarray*}
and  $$\bar P= -\frac{r \chi_0\eta_0^3}{2L(1-2a\chi_0 )}\int \cos
3u\, dx,$$ where $u=kx+\varphi (x)$. One can substitute $P(R)$ and
$\bar P$ back into the potential expression and obtain:
\begin{eqnarray} \nonumber
\tilde\Phi =&&\int dx\{ \frac{D}{2}\eta_0^2(\frac{\partial\varphi}
{\partial x})^2
+\frac{8f-r^2\chi_0}{16}\eta_0^6\cos 6u-\\
&&-\frac{r^2a \chi_0^2\eta_0^6}{4L^2(1-2a\chi_0 )}\left(\int \cos
3u\, dx\right)^2\}
\end{eqnarray}
We once more note a non-local character of the coupling in this
potential, which is given by the square of the integral in the
equation.  Such a coupling is induced by the above-introduced
microscopic electric field, and it disappears at $a=0$. Eq.(2)
allows one (via minimization over $\varphi (R)$) to derive the
equation for the temperature evolving of the IC structure. For the
case $a=0$ the solution of the corresponding equation is well
studied,\cite{Dz} and it gives IC domains of equal sizes, which
are increasing (logarithmically diverging) on cooling towards the
lock-in transition. The dielectric susceptibility for such a
behavior was calculated by Nattermann and Trimper, \cite{Natter}
who showed that it diverges at the lock-in transition point.
However for the case $a\neq 0$, eq.(2) significantly differs from
that for $a=0$. It is easy to check that eq.(2) is equivalent to:
\begin{eqnarray}
\nonumber \tilde\Phi =&&\int dx\{
\frac{D}{2}\eta_0^2(\frac{\partial\varphi}
{\partial x})^2 +\frac{f_1}{2}\eta_0^6\cos (6kx+6\varphi )+\\
&&\frac{r_1}{2} P_1\eta_0^3\cos (3kx+ 3\varphi )
+\frac{\chi^{-1}_1}{2} P_1^2 \},
\end{eqnarray}
where $P_1$ is a spatially constant vector and \[\chi_1^{-1}=
\chi_0^{-1}-2a,\;f_1 =f-r^2\chi_0/8,\; r= r\sqrt{2a\chi_0}.\] In
order to check that at $a\neq 0$ eq.(3) should give the same
solutions for the phase $\varphi (R)$ as eq.(2), it is sufficient
to minimize eq.(3) over $P_1$, and, as a result, return to eq.(2).

On the other hand, eq.(3) coincides by form with eq.(1), with $P$
replaced by the spatially constant vector $P_1$.  Just the
potential of eq.(3) with constant $P_1$ was used by Aslanyan
\cite{Aslanyan} (though the approach was not commented and
sufficiently clarified there) for checking the stability of the IC
phase, and it was shown that within the IC structure a phase
transition should take place from the IC to the ferroelectric IC
phase, prior to the lock-in transition on cooling. The spatial
distribution of the phase $u=\varphi +kx$ and polarization $P(R)$
in this ferroelectric phase are those shown in
Fig.1.\cite{Aslanyan}

In other words, for calculation of the dielectric susceptibility
$\chi^{-1}=\partial^2\tilde\Phi /\partial P^2$ at $ P=0$ one
should minimize potential (1) with respect to $\varphi$, and find
the solution of the equation:
$$D\frac{\partial^2 u}{\partial x^2}+3f\eta_0^4\sin 6u +
\frac{3r}{2}P\eta_0\sin 3u =0.$$ In the paper ref.[4] the
corresponding solution was found for the case $P=const$ in this
equation (which corresponds to minimization of eq. (3)), and in
ref.[3] that was obtained for a spatially dependent $P(R)$.

So, we showed that thermodynamic potential (1), used for
calculation of the dielectric susceptibility, should be
supplemented by the term, contributed from the domain-induced
Lorentz local electric field. As a result, one can see that
calculations of the dielectric susceptibility by Nattermann and
Trimper \cite{Natter} give correct result only for a pair of the
adjacent domains, which is separated from the crystal (i.e. for a
single dipole). For a large number of pairs of domains,
periodically disposed in space as in Fig.1 (in the real IC phase),
the coupling between the dipoles should be taken into account,
which give phase transition to the ferroelectric IC phase with
non-equisize IC domains. This coupling gives contribution of a
negative sign to the potential, and makes ferroelectric IC phase
more favorable compared to the equisize domain IC structure on
approaching the lock-in transition.

As an experimental confirmation of the presented theory may be
pointed out the second-harmonic generation in the IC phases of the
$\bf\rm K_2SeO_4$ \cite{Yes}, $\bf\rm (NH_4)_2BeF_4$ \cite{Alex}
and, under some assumptions, of the quartz \cite{Dolino} crystals,
which is discussed in ref.[4], and the present paper should be
viewed as an additional foundation of its evidence. It is also
worth to note that the developed theory allows one to explain such
a surprising observations as a significant drop in the $\bf\rm
K_2SeO_4$ elastic constant $c_{55}$ (which corresponds to the
transversal $u_z$ sound propagating in the $x$ direction) near the
lock-in transition point \cite{c55}, and observation of a
depolarized, overdamped Raman scattering in the $z(yz)x$ geometry
\cite{Unruh} in the IC phase. These observations unambiguously
follow from the presented theory, and will be published elsewhere.

\end{document}